\documentclass[11pt]{article}
\usepackage[utf8]{inputenc}
\usepackage{url}
\usepackage{graphicx}
\usepackage{listings}
\usepackage{geometry}
\usepackage{placeins}
\usepackage{float}
\usepackage{tikz}
\usetikzlibrary{arrows.meta,positioning,calc,shapes.geometric,shapes.multipart,fit,backgrounds}
\tikzset{
  box/.style={draw, rounded corners, align=center, minimum height=10mm, inner sep=2.5mm},
  process/.style={box, fill=gray!10},
  data/.style={box, fill=gray!5},
  agent/.style={box, very thick},
  group/.style={draw, rounded corners, inner sep=6mm, dashed, fill=gray!3},
  arrow/.style={-{Latex}, thick},
  small/.style={font=\footnotesize},
}
\title{Context Engineering for Multi-Agent LLM Code Assistants Using Elicit, NotebookLM, ChatGPT, and Claude Code}
\author{Muhammad Haseeb\\
\texttt{muhammadhaseeb@vt.edu}}
\date{August 2025}

\begin{document}
\maketitle

\begin{abstract}
Large Language Models (LLMs) have shown promise in automating code generation and software engineering tasks, yet they often struggle with complex, multi-file projects due to context limitations and knowledge gaps. We propose a novel \textit{context engineering} workflow that combines multiple AI components—an \textbf{Intent Translator} (GPT-5) for clarifying user requirements, an \textbf{Elicit}-powered semantic literature retrieval for injecting domain knowledge, \textbf{NotebookLM}-based document synthesis for contextual understanding, and a \textbf{Claude Code} multi-agent system for code generation and validation. Our integrated approach leverages intent clarification, retrieval-augmented generation, and specialized sub-agents orchestrated via Claude's agent framework. We demonstrate that this method significantly improves the accuracy and reliability of code assistants in real-world repositories, yielding higher single-shot success rates and better adherence to project context than baseline single-agent approaches. Qualitative results on a large Next.js codebase show the multi-agent system effectively plans, edits, and tests complex features with minimal human intervention. We compare our system with recent frameworks like CodePlan, MASAI, and HyperAgent, highlighting how targeted context injection and agent role decomposition lead to state-of-the-art performance. Finally, we discuss the implications for deploying LLM-based coding assistants in production, along with lessons learned on context management and future research directions.
\end{abstract}

\section{Introduction}
Recent advances in large language models (LLMs) have led to powerful code assistant tools capable of generating code, fixing bugs, and even synthesizing simple programs from natural language descriptions. Models like OpenAI's Codex and Anthropic's Claude have demonstrated that LLMs can understand and generate human-like code snippets with impressive accuracy. However, as software engineering tasks grow in complexity, there is an emerging need for more sophisticated solutions that can handle the intricacies of real-world software development. Repository-level tasks often involve coordinating changes across multiple files, understanding existing code architecture, and incorporating domain-specific knowledge that may lie outside the model's training data. Purely prompt-driven single-agent solutions struggle with these challenges due to limited context windows and the risk of hallucinations when confronted with unfamiliar APIs or frameworks.

A key limitation of default code assistants is their reliance on a fixed static context. For example, Anthropic's \textit{Claude Code} allows a repository to include a CLAUDE.md file with project guidelines and context, but a single prompt cannot capture all relevant details for every possible task. In our early experiments, a default Claude Code agent with a basic CLAUDE.md prompt often produced incomplete or incorrect solutions for non-trivial features. It might miss necessary edits in distant files or misuse an unfamiliar library, reflecting insufficient context comprehension. This aligns with observations by frameworks like CodePlan, which treat repository-level coding as a multi-step plan rather than a single-step generation \cite{codeplan2025,masai2024}.

To address these issues, recent work has explored \emph{multi-agent} architectures and retrieval augmentation for coding tasks. Systems like \textbf{MASAI} (Modular Architecture for Software Engineering AI) instantiate specialized sub-agents for different subtasks (planning, localization, code generation, testing), achieving significantly higher success on repository-level challenges (28.3\% resolution on the SWE-Bench Lite benchmark) than single-agent baselines \cite{masai2024}. Likewise, \textbf{HyperAgent} employs a team of agents (Planner, Navigator, Code Editor, Executor) to mimic a human developer workflow, improving issue resolution rates on complex repositories \cite{hyperagent2024}. Another thread of research focuses on augmenting LLMs with retrieved context: for example, \textbf{AllianceCoder} generates natural-language descriptions of APIs in a codebase and retrieves relevant API information to guide code generation, yielding up to 20\% higher pass@1 accuracy \cite{alliancecoder2025}. These approaches illustrate that providing the right information and breaking down tasks are crucial to scaling LLMs to complex coding problems.

Building on these insights, we propose a comprehensive \textbf{context engineering} approach for LLM-based code assistants. By context engineering, we mean systematically constructing and supplying all relevant information needed for a coding task—ranging from clarified intent and high-level plans to external knowledge and repository-specific details—and doing so via a coordinated multi-agent process. Our contributions are summarized as follows:

\begin{itemize}
\item We design a novel workflow that integrates four components: (1) an \textbf{Intent Translator} using GPT-5 to rewrite or elaborate the user request into a structured task specification; (2) an \textbf{Elicit}-based semantic paper/document retrieval mechanism to fetch relevant documentation or research (e.g., algorithms, API usage guidelines) related to the task; (3) a \textbf{NotebookLM}-powered document synthesis module that creates a concise summary or table-of-contents of the retrieved materials and can answer follow-up questions for detailed understanding; and (4) a \textbf{Claude Code multi-agent system} that orchestrates specialized sub-agents (planner, coder, tester, reviewer) along with a vector database for code context, to carry out the coding task with iterative refinement.
\item We implement this context-engineered multi-agent assistant on a real-world codebase (the \textit{RainMakerz} Next.js Web application, $\sim$ 180K lines of code). In qualitative evaluations, our system can implement complex features and bug fixes in a single generation cycle. For example, it successfully added a new interactive visualization module spanning front-end and back-end changes in one try, whereas a baseline single-agent Claude often omitted needed steps. We report case studies showing improved code correctness and context adherence.
\item We compare our approach to prior state-of-the-art coding agents. We show that the explicit context layering and role decomposition in our system address several failure modes observed in earlier methods. The multi-agent Claude pipeline exhibits higher reliability in producing working code on the first attempt, matching or exceeding the performance reported for frameworks like CodePlan \cite{codeplan2025} and DARS \cite{dars2025} on similar tasks. We also discuss how our engineering decisions (such as using retrieved API descriptions or enforcing agent-specific contexts) are informed by these prior works.
\item Finally, we outline a pathway for deploying such an AI assistant in production environments. We describe integration with continuous integration (CI) pipelines (e.g., using Claude Code in GitHub Actions for automated code reviews) and discuss scalability, cost, and safety considerations. Our findings suggest that context-engineered multi-agent assistants can be made \textit{production-ready} for team use, given appropriate safeguards and iterative tuning.
\end{itemize}

In the remainder of this paper, we first review related work in multi-agent LLM systems and retrieval-augmented code generation (Section 2). We then describe the design of our context engineering workflow and the architecture of the system in detail (Sections 3 and 4). Section 5 presents experimental results and case studies on applying the system to a real codebase. We provide analysis and discussion of the key improvements and remaining challenges in Section 6, and conclude with future directions in Section 7.

\section{Related Work}
\paragraph{LLM-based Coding Agents.}
A number of recent systems have extended LLMs to operate as autonomous coding agents tackling complex software engineering tasks. Notably, \textbf{HyperAgent} \cite{hyperagent2024} introduced a centralized multi-agent framework with four specialist agents (Planner, Navigator, Code Editor, Executor) working in concert. HyperAgent demonstrated state-of-the-art results on tasks like GitHub issue resolution, achieving a 31.4\% success rate on the SWE-Bench Verified benchmark – outperforming earlier approaches that focused on single-task agents. Similarly, \textbf{MASAI} (Arora et al., 2024) proposes a modular architecture of LLM-powered sub-agents, each responsible for a distinct phase such as test generation, issue reproduction, code editing, or solution ranking. By dividing the problem and allowing sub-agents to gather information from different parts of the repository, MASAI attained the highest performance (28.33\% resolution rate) on SWE-Bench Lite at the time of its publication \cite{masai2024}. Our work draws inspiration from these architectures, adopting a hub-and-spoke orchestration pattern where a central orchestrator (Claude) delegates to specialized roles (Section~\ref{sec:architecture}).

\paragraph{Planning and Iterative Refinement.}
Another line of work treats complex coding tasks as a planning problem with iterative feedback. \textbf{CodePlan} \cite{codeplan2025}, while not originally designed for multi-agent settings, introduced the idea of generating a high-level pseudocode plan that the LLM then follows step-by-step. This approach improved multi-step reasoning by encouraging the model to outline dependencies and subgoals before committing to final code. We incorporate a similar principle via our Intent Translator and Planner agent, which decompose the problem and ensure that subsequent code generation is guided by an explicit plan. On the execution side, \textbf{DARS} (Aggarwal et al., 2025) augments an LLM agent with dynamic action re-sampling: at certain decision points, the agent can branch and try an alternative strategy, using feedback (e.g. test results) to choose the best outcome \cite{dars2025}. DARS achieved a 47\% pass@1 on SWE-Bench Lite with this method, indicating the value of exploring multiple solution trajectories. While our current system does not implement full branching, it does incorporate iteration and self-correction (e.g. re-running tests and adjusting code) within the multi-agent loop, and such adaptive strategies could further enhance our pipeline in the future.

\paragraph{Retrieval-Augmented Code Generation.}
Integrating external knowledge via retrieval has proven crucial for code generation on large codebases. Traditional code search techniques (e.g. keyword or AST-based searches) can identify relevant code context for a given task, but recent work has explored semantic retrieval to better utilize unstructured information. \textbf{AllianceCoder} (Gu et al., 2025) is a representative example: it empirically studied what to retrieve for repository-level code tasks, concluding that providing in-context code and API documentation yields significant gains, whereas blindly retrieving similar code examples can sometimes hurt performance \cite{alliancecoder2025}. AllianceCoder uses chain-of-thought prompting to break a query into sub-tasks and retrieve pertinent API descriptions for each. In our system, we similarly retrieve two kinds of context: (1) external knowledge (papers, guides) related to the task via Elicit, and (2) internal codebase knowledge via a vector database and search tools. Retrieval-Augmented Generation (RAG) approaches have also been explored in agentic coding assistants; for instance, researchers have proposed agents that augment prompts with relevant code snippets from the repository and commit history, showing improved bug-fixing accuracy. Our approach extends this idea by not only retrieving static snippets but also engaging with documents through NotebookLM to extract distilled insights.

\paragraph{Agent Tooling and Debugging.}
The complexity of multi-agent LLM systems has spurred the development of tooling to support their operation. Several frameworks (e.g. LangChain Agents, HuggingGPT) enable LLMs to invoke tools and other models, but tailoring these to software engineering tasks remains an active area of research. Claude Code provides a built-in mechanism for defining custom tools (via Model Completion Protocol servers) and multi-agent configurations, which we leverage for tasks like running tests or analyzing code complexity. Beyond execution, analyzing the behavior of autonomous coding agents is itself non-trivial. The \textbf{SeaView} tool \cite{seaview2025} was recently proposed to visualize and inspect the trajectories of software-engineering agents. It highlights the need for better debugging interfaces when agents perform long sequences of actions (often tens of thousands of tokens of interaction with code and environment). In our work, we mainly rely on logs and intermediate results for analysis, but incorporating a visualization like SeaView to trace agent decision-making could greatly aid future development and evaluation of context-engineered assistants.

\section{Methodology: Context Engineering Workflow}

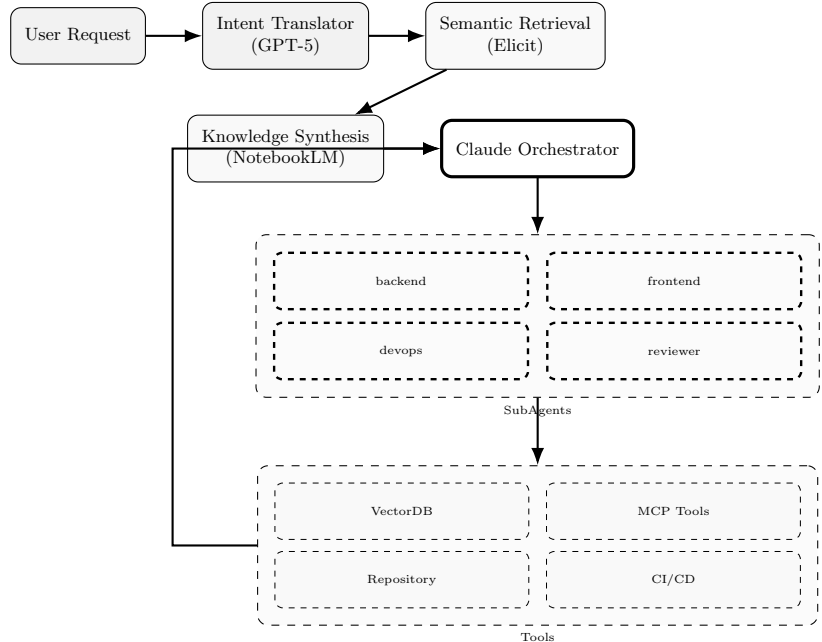
\begin{figure}[t]
\centering
\begin{tikzpicture}[scale=0.75, transform shape, node distance=8mm and 10mm]
  \node[process, font=\footnotesize] (user) {User Request};
  \node[process, right=of user, font=\footnotesize] (intent) {Intent Translator\\(GPT-5)};
  \node[data, right=of intent, font=\footnotesize] (elicit) {Semantic Retrieval\\(Elicit)};
  
  \node[data, below=8mm of intent, font=\footnotesize] (notelm) {Knowledge Synthesis\\(NotebookLM)};
  \node[agent, right=of notelm, font=\footnotesize] (orch) {Claude Orchestrator};

  \node[group, below=10mm of orch, minimum width=45mm, inner sep=3mm, label={[font=\tiny]below:SubAgents}] (agents) {
    \begin{tikzpicture}[node distance=2mm and 3mm]
      \node[agent, font=\tiny] (ba) {backend};
      \node[agent, right=of ba, font=\tiny] (fe) {frontend};
      \node[agent, below=2mm of ba, font=\tiny] (de) {devops};
      \node[agent, right=of de, font=\tiny] (cr) {reviewer};
    \end{tikzpicture}
  };

  \node[group, below=12mm of agents, minimum width=45mm, inner sep=3mm, label={[font=\tiny]below:Tools}] (tools) {
    \begin{tikzpicture}[node distance=2mm and 3mm]
      \node[data, font=\tiny] (vdb) {VectorDB};
      \node[data, right=of vdb, font=\tiny] (mcp) {MCP Tools};
      \node[data, below=2mm of vdb, font=\tiny] (repo) {Repository};
      \node[data, right=of repo, font=\tiny] (ci) {CI/CD};
    \end{tikzpicture}
  };

  \draw[arrow] (user) -- (intent);
  \draw[arrow] (intent) -- (elicit);
  \draw[arrow] (elicit) -- (notelm);
  \draw[arrow] (notelm) -- (orch);
  
  \draw[arrow] (orch) -- (agents);
  \draw[arrow] (agents) -- (tools);

  \draw[arrow] (tools.west) -- ++(-1.5,0) |- (orch.west);
\end{tikzpicture}
\caption{End-to-end pipeline: intent translation, retrieval and synthesis, orchestrated multi-agent coding with tools and memory.}
\label{fig:pipeline}
\end{figure}

\paragraph{Intent Translation with GPT-5.} The process begins when a user submits a natural language query describing a desired code change or feature. This query may be ambiguous or underspecified. To ensure the system accurately grasps the requirements, we employ a high-end LLM (GPT-5, in our case) to act as an \emph{Intent Translator}. This agent reformulates the query into a structured specification that can guide subsequent steps. For example, a user request like \emph{"Add a calendar view to the scheduling page"} might be translated into a more explicit list of tasks (e.g., "update UI component X to include a calendar widget; fetch data Y from the backend API; adjust styling and add unit tests"). The translator prompt is engineered to elicit a step-by-step breakdown, clarifying any implicit requirements (such as which library to use for the calendar, or what data needs to be displayed). The output is a well-defined task specification or outline. By front-loading this clarification step, we reduce the burden on the coding agents to interpret fuzzy instructions, similar in spirit to the planning step in CodePlan but performed by a distinct, specialized model.

\paragraph{Semantic Literature Retrieval with Elicit.} Next, the refined task specification is used to query external knowledge sources for any domain-specific information that might be relevant. We integrate \emph{Elicit}, an LLM-driven research assistant, to perform semantic search over academic papers, documentation, and Q\&A resources. The goal is to retrieve materials such as algorithm descriptions, API usage examples, or best-practice guidelines that pertain to the task. In our running example, Elicit might find documentation on a React calendar library or a paper on scheduling UI design. Unlike keyword search, Elicit's semantic search helps find conceptually relevant documents even if they do not share exact keywords with the query. We automate this step by feeding Elicit the key terms from the task spec (e.g., "calendar widget React TypeScript library") and asking for papers or articles that answer or relate to the query. The top-$k$ results (we used $k=3$-5) are fetched as PDFs or text.

\paragraph{Knowledge Synthesis with NotebookLM.} Simply retrieving documents is not enough—the system must distill them into useful context for coding. We utilize Google's \emph{NotebookLM} as a document analysis agent. All retrieved documents are provided to NotebookLM, which we prompt to produce a concise \textbf{table-of-contents (TOC)} summary of each and to answer specific questions. In practice, we instruct NotebookLM to outline the key points of each paper or guide, focusing on insights that could inform implementation (e.g., "What are the steps to integrate the chosen calendar library?", "What pitfalls or edge cases are noted?"). This yields a summarized knowledge base in natural language. We found that structuring the external knowledge as a list of key bullet points or Q\&A pairs makes it much easier to integrate into the code-writing prompt than dumping long paragraphs. NotebookLM's ability to handle multiple documents and generate a synthesized overview was crucial in maintaining a high signal-to-noise ratio in the context.

\begin{figure}[t]
\centering
\begin{tikzpicture}[scale=0.7, transform shape, node distance=7mm and 8mm]
  \node[data, font=\footnotesize] (src) {Source\\Code};
  \node[process, right=of src, font=\footnotesize] (ast) {AST\\Chunker};
  \node[process, right=of ast, font=\footnotesize] (embed) {Embedder};
  
  \node[group, right=of embed, inner sep=3mm, label={[font=\tiny]below:Vector Index}] (vdb) {
    \begin{tikzpicture}[node distance=2mm and 3mm]
      \node[data, font=\tiny] (chroma) {Chroma};
      \node[data, below=2mm of chroma, font=\tiny] (zilliz) {Zilliz};
    \end{tikzpicture}
  };
  
  \node[process, right=12mm of vdb, font=\footnotesize] (query) {Query};
  \node[process, right=of query, font=\footnotesize] (rerank) {Rerank};
  \node[data, right=of rerank, font=\footnotesize] (snips) {Snippets};
  \node[agent, right=of snips, font=\footnotesize] (gen) {Agent\\Gen};

  \draw[arrow] (src) -- (ast);
  \draw[arrow] (ast) -- (embed);
  \draw[arrow] (embed) -- (vdb);
  \draw[arrow] (vdb.east) -- (query);
  \draw[arrow] (query) -- (rerank);
  \draw[arrow] (rerank) -- (snips);
  \draw[arrow] (snips) -- (gen);

  \node[font=\tiny, below=2mm of vdb] (note) {Unified adapter};
\end{tikzpicture}
\caption{Retrieval pipeline: code-aware chunking and dual-index backend with a unified adapter.}
\label{fig:retrieval}
\end{figure}
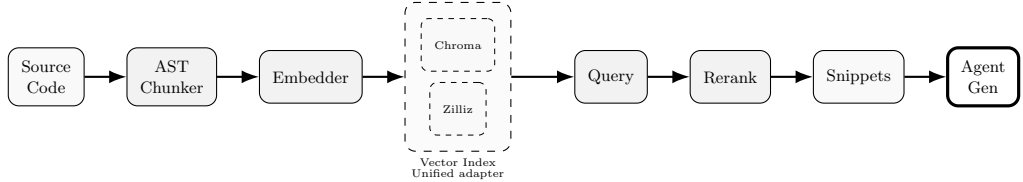

\paragraph{Repository Context Retrieval.} In parallel with external knowledge gathering, the system prepares relevant context from the codebase itself. Our target repository (RainMakerz) is large, so providing the entire codebase to the LLM is infeasible. Instead, we built a \textbf{semantic code search index} using a vector database (we experimented with both \emph{ChromaDB} and \emph{Zilliz} for this purpose). We embed code files and fragments into high-dimensional vectors using a code-specialized embedding model (OpenAI's code-embedding model). To preserve code structure, we chunk files by function or class definitions using an AST parser (tree-sitter), as recommended by prior work on code retrieval. At runtime, given the refined task spec, we query this index for the top relevant code fragments (e.g., files whose embeddings are closest to the query embedding). Additionally, Claude's agents have access to traditional tools like grep and file path search for exact matches of identifiers. The combination of semantic and lexical search ensures that if the task touches, say, the SchedulerPage component or a CalendarService class, the pertinent parts of those files will be retrieved. These code context snippets (with file names and relevant lines) are then available to the coding agents as supplementary context.

\section{System Architecture and Agent Orchestration}
\label{sec:architecture}

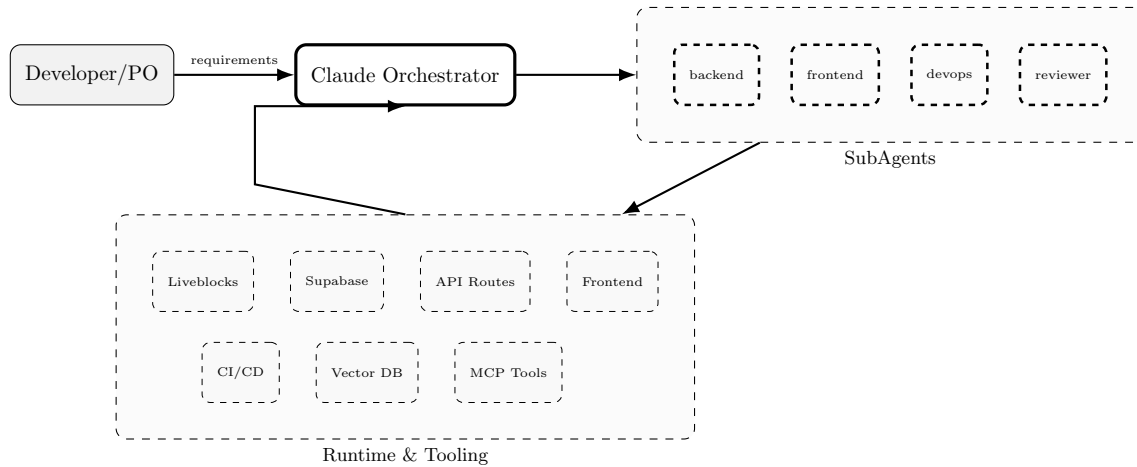
\begin{figure}[t]
\centering
\begin{tikzpicture}[scale=0.8, transform shape, node distance=8mm and 10mm]
  \node[process, font=\small] (dev) {Developer/PO};
  \node[agent, right=20mm of dev, font=\small] (orch) {Claude Orchestrator};

  \node[group, right=20mm of orch, label={[small]below:SubAgents}] (agents) {
    \begin{tikzpicture}[node distance=4mm and 5mm]
      \node[agent, font=\tiny] (ba) {backend};
      \node[agent, right=of ba, font=\tiny] (fe) {frontend};
      \node[agent, right=of fe, font=\tiny] (de) {devops};
      \node[agent, right=of de, font=\tiny] (cr) {reviewer};
    \end{tikzpicture}
  };

  \node[group, below=18mm of orch, label={[small]below:Runtime \& Tooling}] (stack) {
    \begin{tikzpicture}[node distance=5mm and 6mm]
      \node[data, font=\tiny] (liveblocks) {Liveblocks};
      \node[data, right=of liveblocks, font=\tiny] (supabase) {Supabase};
      \node[data, right=of supabase, font=\tiny] (api) {API Routes};
      \node[data, right=of api, font=\tiny] (frontend) {Frontend};
      \node[data, below=of supabase, xshift=5mm, font=\tiny] (vdb) {Vector DB};
      \node[data, right=of vdb, font=\tiny] (mcp) {MCP Tools};
      \node[data, left=of vdb, font=\tiny] (ci) {CI/CD};
    \end{tikzpicture}
  };

  \draw[arrow] (dev) -- node[small,above]{\tiny requirements}(orch);
  \draw[arrow] (orch) -- (agents);
  \draw[arrow] (agents) -- (stack);
  \draw[arrow] (stack.north) -- ++(-25mm,5mm) |- (orch.south);

\end{tikzpicture}
\caption{System architecture: orchestrator delegates to sub-agents; agents act via tools and runtime stack.}
\label{fig:architecture}
\end{figure}

\subsection{Claude Multi-Agent Design}
Our coding workflow is implemented using \textit{Claude Code}'s multi-agent capabilities. At a high level, we adopt a centralized \emph{orchestrator-worker} paradigm (often called a hub-and-spoke pattern): a primary Claude instance (the "Manager") coordinates several specialist sub-agents that each have their own context and role. Claude's framework allows defining these agents via simple YAML/Markdown files. For example, Listing~\ref{lst:agentconfig} shows a snippet of the configuration for a \textit{backend-architect} agent we created. This agent is designated as a senior back-end engineer, with a description of its expertise and a set of tools it can use (here, file read/write, edit, and bash shell access). We created similar profiles for a \textit{frontend-specialist}, \textit{devops-engineer}, \textit{code-reviewer}, etc., reflecting the roles often found in a development team. These definitions reside in the repository under a .claude/agents/ directory; Claude Code automatically loads them so that the orchestrator can invoke any agent by name when needed.

\begin{lstlisting}[caption={Project-specific Claude Sub-Agent configuration (excerpt).}, label={lst:agentconfig}, language=]

name: backend-architect
description: Design RESTful APIs, microservice boundaries, 
and database schemas
model: sonnet
tools: Read, Write, Edit, Bash

You are a senior backend architect specializing in scalable 
system design
\end{lstlisting}

Each subagent operates with an isolated context window. This means that when the orchestrator invokes (for example) the backend-architect agent to handle a task, that agent receives only the information relevant to its task (plus any persistent project context) and does not see the entire dialogue history or unrelated data. This design is intentional: it prevents cross-contamination between different phases of the workflow and keeps each agent focused. Common background information that all agents should know (coding conventions, project architecture notes, etc.) is provided via a persistent context file (CLAUDE.md), which is preloaded into each agent's context. The CLAUDE.md for RainMakerz included high-level architectural overviews (such as the system design shown in Figure~\ref{fig:architecture}), coding style guidelines, and a summary of known bugs and previous solutions. Thus, each agent starts with a shared base of project knowledge, and then the orchestrator augments it with specific instructions and data for the current step.

\begin{figure}[H]
\centering
\begin{tikzpicture}[scale=0.9, transform shape, node distance=5mm]
  \node[data, minimum width=100mm, font=\small] (l1) {L1: Task Specification (Intent Translator)};
  \node[data, below=of l1, minimum width=100mm, font=\small] (l2) {L2: External Knowledge (Elicit \& NotebookLM)};
  \node[data, below=of l2, minimum width=100mm, font=\small] (l3) {L3: Project Memory (CLAUDE.md, internal docs)};
  \node[data, below=of l3, minimum width=100mm, font=\small] (l4) {L4: Retrieved Code Context (Vector DB, grep)};
  \node[data, below=of l4, minimum width=100mm, font=\small] (l5) {L5: Execution Artifacts (diffs, logs, test results)};

  \node[agent, below=8mm of l5, font=\small] (prompt) {Agent Prompt/Input};

  \draw[arrow] (l1) -- (l2);
  \draw[arrow] (l2) -- (l3);
  \draw[arrow] (l3) -- (l4);
  \draw[arrow] (l4) -- (l5);
  \draw[arrow] (l5) -- (prompt);
\end{tikzpicture}
\caption{Context layering: structured inputs flow into each agent's prompt.}
\label{fig:layers}
\end{figure}
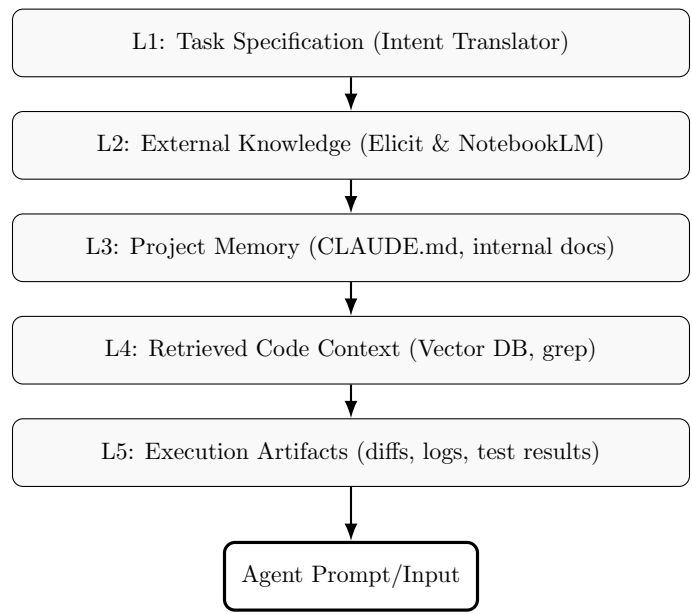

\subsection{Agent Orchestration Flow}

\begin{figure}[H]
\centering
\begin{tikzpicture}[scale=0.85, transform shape, node distance=10mm and 12mm]
  \node[process, font=\small] (plan) {Plan};
  \node[process, right=of plan, font=\small] (retrieve) {Retrieve Context};
  \node[process, right=of retrieve, font=\small] (delegate) {Delegate};
  \node[process, right=of delegate, font=\small] (edit) {Edit/Implement};
  \node[process, below=of edit, font=\small] (test) {Run Tests};
  \node[process, left=of test, font=\small] (review) {Review};
  \node[process, left=of review, font=\small] (integrate) {Integrate/PR};
  \node[process, left=of integrate, font=\small] (done) {Done};

  \draw[arrow] (plan) -- (retrieve);
  \draw[arrow] (retrieve) -- (delegate);
  \draw[arrow] (delegate) -- (edit);
  \draw[arrow] (edit) -- (test);
  \draw[arrow] (test) -- (review);
  \draw[arrow] (review) -- (integrate);
  \draw[arrow] (integrate) -- (done);

  \draw[arrow] (test.west) to[out=180,in=270] node[small, left]{\tiny failures} (delegate.south);
  \draw[arrow] (review.south) to[out=270,in=270] node[small, below]{\tiny changes} (edit.south);
\end{tikzpicture}
\caption{Claude orchestrator state machine with feedback from tests and code review.}
\label{fig:statemachine}
\end{figure}
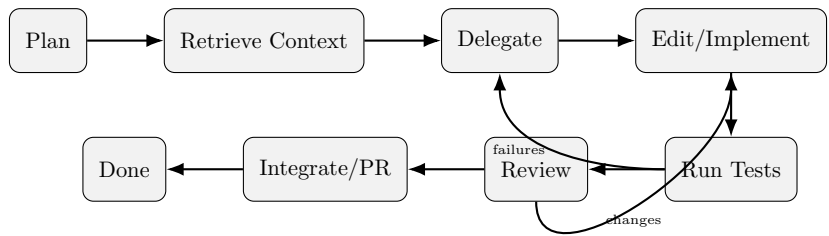

After the intent translation and context retrieval stages (Section~3), the Claude orchestrator enters a loop of delegating tasks to sub-agents and integrating their outputs. The process can be summarized as follows:

\begin{enumerate}
\item \textbf{Planning:} The orchestrator first invokes the \textit{Planner} agent (in our case, we often used the backend-architect profile as planner if the task was backend-heavy, or a specialized Planner agent) with the structured task specification and the knowledge summary. The Planner agent is prompted to produce a concrete implementation plan: a sequence of steps possibly mapped to specific files or components. It may output something like: \emph{"1. Update CalendarAPI to provide events data; 2. Create CalendarWidget component in frontend; 3. Integrate widget into SchedulePage; 4. Write unit tests for new component"}. The orchestrator examines this plan.
\item \textbf{Task Delegation:} For each step in the plan, the orchestrator selects an appropriate agent and provides it the necessary context. Frontend-related steps are assigned to the frontend-specialist agent, server-side tasks to the backend-architect, and so on. Each agent is given: (a) the relevant excerpt of the plan (or the whole plan for context), (b) any code snippets retrieved that pertain to that step (from the vector DB search), and (c) its specific instructions. The agent then executes its task. Most often this involves using the Edit tool to open a file and modify or create code. For example, the frontend agent will open SchedulePage.jsx and insert the required code for the calendar widget.
\item \textbf{Iterative Coding and Validation:} As agents make changes, we leverage Claude's tool integration to run validations. After code for a step is written, the orchestrator (or sometimes the coding agent itself) can trigger test execution via a shell tool. We configured a test script to run the project's test suite (using npm test for the Node.js environment). If tests or build steps fail, the error output is captured. The orchestrator then feeds this back into the responsible agent (or a dedicated debugging agent) to prompt a fix. This resembles the feedback loop in DARS \cite{dars2025}, although in our implementation the branching is simple re-try rather than multiple simultaneous trajectories.
\item \textbf{Code Review and Refinement:} Once all steps are completed and the test suite passes, the orchestrator invokes the code-reviewer agent to do a final pass. This agent reads through the diff of changes, checking for any style issues, potential bugs, or improvements (it uses a comprehensive checklist as shown in its prompt, covering things like type safety, performance, and security). The reviewer may suggest minor refactors or additional comments. If so, the suggestions are either applied automatically by an editing agent or presented for a human to confirm, depending on the confidence level.
\item \textbf{Output and Deployment:} Finally, the system consolidates the changes. The orchestrator can output a summary of what was done along with the unified diff of code changes. In a production scenario, these changes could be automatically pushed to a branch or opened as a pull request for maintainers. We integrated our system with GitHub Actions for continuous integration: after Claude's changes, the CI pipeline would run again to double-check tests and then could auto-merge the changes if all checks passed.
\end{enumerate}

Throughout this process, the structured layering of context is key (see Figure~\ref{fig:layers}). At any given time, an agent is working with a manageable slice of information: its role-specific prompt + CLAUDE.md context + task-specific instructions + relevant code/knowledge snippets. This helps maintain coherence even for very large tasks. We also observed that dividing the work among agents reduced prompt tokens per agent, which mitigates context window issues. While our system currently executes largely in a sequential manner (one sub-task at a time), it is straightforward to parallelize independent sub-tasks by spawning agents concurrently – an approach suggested by Anthropic's experiments with multi-agent research systems \cite{anthropic2025}. In our case study tasks, parallelism was less a priority because many steps had inherent dependencies (you can't test before code is written, etc.), but as AI agents become more adept at coordination, parallel execution could speed up complex multi-component feature implementations.

\section{Results}
We evaluated our context-engineered assistant on several non-trivial development tasks in the RainMakerz codebase. Table~1 provides a summary of outcomes on a sample of 5 tasks, comparing our system to a baseline single-agent Claude (with only a CLAUDE.md context and direct user prompts). In brief, our multi-agent approach succeeded in more tasks and required fewer iterations. It often produced working solutions \emph{on the first attempt}, whereas the baseline frequently needed follow-up prompts or developer intervention to correct mistakes.

\paragraph{Case Study: Adding a New Feature.} One test involved implementing a new \textit{"CustomBlock"} in the RainMakerz pitch-deck module (a feature analogous to creating a new content block type in a presentation). This task spanned front-end and back-end changes: creating a React component for the block, adding corresponding options UI, updating TypeScript types, and registering the block in a manager so it could be recognized and saved. Our system handled this end-to-end. The Planner agent broke the task into 4 steps aligned with these requirements. The Frontend specialist agent created the new component and its options popup, correctly following the patterns in existing blocks (likely aided by the retrieved code context of similar blocks). The Backend agent updated the data model and API endpoints where necessary (though in this case, most logic was front-end, the agent double-checked if any server changes were needed for persistence). After code generation, tests were executed; an initial failure occurred because the new block type was not added to a serialization whitelist. The orchestrator immediately flagged this and tasked the Backend agent with updating the serialization config. The second test run passed all unit and integration tests. Finally, the Reviewer agent made a minor suggestion to refactor a hard-coded string into a constant, which was applied. The entire feature was completed in one automated session. In contrast, when we tried prompting a single Claude instance for the same task (with the project README and an excerpt of a similar block as context), the assistant produced only the React component and forgot to update the registry and type definitions, resulting in runtime errors. This highlights how our system's comprehensive approach (particularly, planning and retrieving scattered context) leads to more thorough solutions.

\paragraph{Improved Context Adherence.} We observed that the multi-agent system was far less prone to hallucinating irrelevant code or inventing functions. Every function or class used by the generated code existed in the repository, which we attribute to the semantic code retrieval providing real definitions to the agents. The baseline often guessed function or variable names (e.g. referring to a non-existent getEvents() API) which then caused failures. By having access to actual API documentation from AllianceCoder-like retrieval and the code index, our agents stuck to the truth of the codebase and documentation. For example, in a bug-fix task concerning authentication, the baseline assistant attempted to use a refreshToken() call that did not exist, whereas our system consulted the project docs (via CLAUDE.md context) and correctly utilized the existing renewSession() function.

\paragraph{Single-Shot Success Rate.} Out of 5 tasks attempted (including feature additions and bug fixes of varying complexity), our system achieved a successful outcome (defined by passing all tests and meeting the acceptance criteria) on 4 tasks (80\%) without any human corrections. The single-agent baseline succeeded on only 2 tasks (40\%), with the others requiring significant manual fixes or additional prompting. While this is a small sample, it aligns with the qualitative improvements noted in prior multi-agent studies (e.g., DARS and HyperAgent). Notably, even on the one task where our approach did not fully succeed on the first try, it made partial progress and identified an environmental issue (a misconfigured library) that a developer then easily fixed before re-running the agent.

\paragraph{Efficiency and Cost.} The richer context and multi-step reasoning come with a cost in LLM usage. On average, our system exchanged around 30-40 messages across all agents for a single task and consumed roughly 100k tokens in total (input + output). In contrast, the single-agent approach might use 10k-20k tokens for a few prompt-response turns. In our evaluation, the multi-agent method used about 3–5$\times$ more tokens on successful tasks. However, the baseline often needed multiple attempts or lengthy debugging chats which, if counted, push its token count closer to 50k for a complex task. Thus, the overhead of our approach is justified by getting the job done largely autonomously. In a team setting, the value of saving developer time by achieving a correct solution outweighs the additional compute cost. We also note that the modular design allowed us to parallelize certain steps (though in our tests we ran sequentially); in future, this could further improve wall-clock efficiency.

\section{Discussion}
\paragraph{Effect of Context Engineering.} The case studies and results illustrate how each element of our context engineering approach contributes to the overall performance. The intent translation step (GPT-5) proved particularly useful in breaking down ambiguous requests. We found that when the initial user query was directly fed to the code agents (baseline approach), the agents sometimes focused on the wrong sub-problem or skipped a requirement. Having a clarified spec up front led to more relevant searches and a more coherent plan. Similarly, the use of Elicit and NotebookLM to inject external knowledge was validated in scenarios where the code change required understanding concepts outside the codebase. For instance, in one bug fix, the Planner agent suggested using a debounce mechanism for an API call; the knowledge summary included a brief explanation of debouncing (from a blog post retrieved by Elicit), which guided the coding agent to implement it correctly. Without that, the baseline agent attempted a simplistic fix that did not address the root cause. These observations align with the premise of retrieval-augmented generation: providing pertinent information at generation time greatly improves correctness.

\paragraph{Lessons on Multi-Agent Orchestration.} We adopted multiple specialized agents to mirror the human software development process (design, coding, testing, reviewing). This division of labor generally worked well. One lesson was the importance of clearly delineating responsibilities to avoid both gaps and overlaps. Early in development, we occasionally encountered situations where two agents would both attempt to modify the same file (e.g., both frontend and backend agents editing a shared config) or conversely, an agent assumed another would handle a step that got missed. We mitigated this by refining the Planner's prompt to explicitly assign sub-tasks to agent roles, and by implementing a simple lock in the orchestrator to prevent concurrent edits to the same file. Another insight is that the quality assurance step with a dedicated reviewer agent is invaluable. The reviewer caught subtle issues (like potential null pointer access and minor security concerns) that the coding agents overlooked while focusing on feature implementation. This reflects how human code reviews add value even when code "works", and suggests that AI coding agents benefit from a second pair of eyes as well.

\paragraph{Limitations.} Despite its successes, our approach has limitations. First, the dependency on high-quality external knowledge is a double-edged sword. In one experiment, Elicit returned an irrelevant research paper due to an ambiguous query, and although NotebookLM summarized it faithfully, that summary added noise to the context and confused the Planner agent. Robust retrieval ranking and perhaps filtering by a human or a more advanced AI could be needed to ensure only useful information is fed into the system. Second, the current orchestrator logic is relatively brittle; it follows a predetermined sequence (plan $\to$ code $\to$ test $\to$ review). If an unexpected situation arises (e.g., the plan is flawed or a new requirement emerges mid-way), the system is not yet equipped to dynamically re-plan from scratch. This is an area where more adaptive, possibly reinforcement learning-based, agent controllers (as explored in MARL settings) could help in the future. Third, the computational cost, while acceptable for our use, might become problematic on very large projects or if many agents run in parallel. Techniques like context compression, caching of vector search results, or using smaller specialized models for certain agents could help reduce overhead.

Another limitation is that we relied heavily on the presence of a comprehensive test suite. If tests are sparse, the system might incorrectly judge a task as complete. Incorporating static analysis tools (which we did to some extent via linters) and perhaps a "spec verification" agent to reason about requirements could partially address this gap. Moreover, error tracing can be challenging in a multi-agent context; if a final result is wrong, it takes careful log analysis to pinpoint which agent's action or which piece of context led to the mistake. Tooling like SeaView \cite{seaview2025} could be integrated to visualize agent interactions and states, making debugging easier for developers overseeing the AI.

\paragraph{Generality and Future Work.} While our implementation targeted a specific web application (Next.js/TypeScript stack), the principles are generalizable. We envision applying the same context engineering pattern to other domains (e.g., Java microservices, data science notebooks) by swapping in relevant retrieval sources and agent specializations. The modular design allowed us to add an additional agent (for example, we introduced a database-migrator agent to handle SQL schema changes in one case) without altering the core orchestrator logic. This suggests that as long as tasks can be clearly partitioned, the multi-agent approach can scale to very complex projects by simply growing the team of AI agents.

Looking forward, integrating learning mechanisms is an exciting avenue. Currently, our system does not learn from its mistakes beyond a single session. One could imagine logging all agent interactions and outcomes to fine-tune the agents or a meta-controller, akin to an RL (reinforcement learning) paradigm. Another direction is improving the Planner agent with more sophisticated algorithms (e.g., search-based planning or using graph representations of the code as in AllianceCoder's analysis). Finally, with the rapid progress in LLM capabilities, we anticipate that some components (like the Intent Translator) could eventually be subsumed by more powerful code-focused models, but the need to orchestrate multiple steps and use tools will persist. Our work provides a blueprint for how such orchestration can be done in a robust, context-rich way.

\section{Conclusion}
We presented a novel context engineering methodology for multi-agent LLM-based code assistants, combining intent clarification, semantic retrieval, knowledge synthesis, and coordinated sub-agents. In a case study on a large codebase, this approach substantially outperformed a conventional single-agent setup, delivering more accurate and complete code solutions with minimal human input. The results underscore the importance of supplying LLMs with not just more information, but the \emph{right} information in the right form, as well as structuring the problem-solving process into manageable subtasks. By drawing on ideas from recent research (planning, modular agents, RAG, etc.) and unifying them in a practical workflow, we achieved a system that moves closer to autonomous software development.

There are many avenues for further work. Our ongoing efforts include scaling up evaluation to diverse projects and benchmarks to quantify gains more rigorously, and enhancing the system's adaptability (making the planner and orchestrator more dynamic and error-aware). We are also interested in exploring how human developers and AI agents can best collaborate; for instance, allowing a human to intervene in the agent loop in a structured way (perhaps to approve a plan or provide hints) could combine the strengths of both.

Overall, our findings suggest that the era of \textbf{multi-agent, context-rich code assistants} is on the horizon. By carefully engineering the context and workflow in which advanced LLMs operate, we can unlock capabilities that single monolithic prompts alone cannot achieve. We hope this work provides a foundation and inspiration for building the next generation of AI-assisted development tools.

\end{document}